\documentclass[aps,prb,twocolumn,superscriptaddress,showpacs]{revtex4}
\usepackage{graphicx}

\begin{document}

\title{Far infrared absorption by acoustic
       phonons in titanium dioxide nanopowders}

\author{Daniel B. Murray}
\email{daniel.murray@ubc.ca}
\author{Caleb H. Netting}
\affiliation{Mathematics, Statistics and Physics Unit,
University of British Columbia - Okanagan, 3333 University Way,
Kelowna, British Columbia, Canada V1V 1V7}

\author{Lucien Saviot}
\email{lucien.saviot@u-bourgogne.fr}
\author{Catherine Pighini}
\author{Nadine Millot}
\author{Daniel Aymes}
\affiliation{Laboratoire de Recherche sur la R\'eactivit\'e des Solides,
UMR 5613 CNRS - Universit\'e de Bourgogne\\
9 avenue A. Savary, BP 47870 - 21078 Dijon - France}

\author{Hsiang-Lin Liu}
\email{hliu@phy.ntnu.edu.tw}
\affiliation{Department of Physics, National Taiwan Normal University,
88, Sec. 4, Ting-Chou Road, Taipei 116, Taiwan}

\date{\today}

\begin{abstract}
We report spectral features of far infrared electromagnetic
radiation absorption in anatase TiO$_2$ nanopowders which we
attribute to absorption by acoustic phonon modes of
nanoparticles.  The frequency of peak excess absorption above
the background level corresponds to the predicted frequency of
the dipolar acoustic phonon from continuum elastic theory.
The intensity of the absorption cannot
be accounted for in a continuum elastic dielectric
description of the nanoparticle material.  Quantum
mechanical scale dependent effects must be considered.
The absorption cross section is estimated from
a simple mechanical phenomenological model.
The results are in plausible agreement with
the absorption being due to a sparse 
layer of charge on the nanoparticle surface.
\end{abstract}

\maketitle

\section{Introduction}

The optical properties of nanoparticles (NP) provide avenues
for material characterization and for the creation of 
nanocomposites for customized photonics applications.
In particular, the mechanical vibrations (i.e. acoustic
phonons) of NP can be observed through low frequency Raman
scattering (LFR) and also through ultrafast laser pump-probe
experiments (UFLPP).  However,
the THz frequency range of these acoustic phonons raises
the question of whether they can absorb electromagnetic
radiation in the far infrared (FIR). This has not been seen
before now. 

Anomalous FIR absorption of small metal NP has long been
predicted due to the discretization of electron energy
levels which would otherwise be in a continuum in bulk
metal.\cite{gorkov65,halperin86}
This has been experimentally investigated in NP of Cu, Al, Sn
and Pb,\cite{tanner75} Al,\cite{granqvist76}
and Ag.\cite{devaty84,lee85,devaty90}

There is significant interest in FIR
absorption of NP molecules (small groups of NPs)
in a strong magnetic field, related to the electronic
states.\cite{helle05b,marlo03,marlohelle05}

However, the FIR properties of insulating NP have been
ignored up to now.  It is the aim of the present paper
to explore FIR absorption of insulating NP with
emphasis on the role of acoustic phonons having
frequencies in the FIR.  It is necessary to proceed with
a simultaneous experimental and theoretical investigation.
Measurements on the FIR absorption of titanium dioxide
NP are presented here, and they indicate the overall
absorption cross section per NP.  Theoretical estimates
are given below of the expected order of magnitude of
the portion of FIR absorption cross section per NP
due to acoustic phonons.

\section{FIR absorption}

Here we review the basic features of FIR absorption.
The general setup of
a FIR absorption experiment is that a FIR source is
arranged in relation to a FIR detector, with the
sample in between.  The detector records the
transmitted intensity $I_T$($\omega$) over a range of
frequency, $\omega$, which went from 15~cm$^{-1}$ to
200~cm$^{-1}$ in our case.  The transmitted intensity
with the sample temporarily removed is $I_{T0}$($\omega$).  The
transmitted intensity ratio attributed to the
effect of the sample is $I_T(\omega)/I_{T0}(\omega)$.  Let $t_{samp}$
be the thickness of the sample in $cm$.  The absorption
coefficient is obtained by
$\alpha(\omega) = (-1/t_{samp}) \ln(I_T/I_{T0})$
where $\ln$ is the natural logarithm.

When the sample consists of $N$ individual NP that each
have absorption cross section $\sigma$ (in square meters)
the total absorbing area is $N\sigma$.  If $A$ is the area
of the aperture in which the sample is placed then
\begin{equation}
\sigma(\omega) = - \frac{A}{N} \ln\left(\frac{I_T}{I_{T0}}\right)
\label{eq_sigma_exp}
\end{equation}

There are two distinct reasons why the incident FIR
radiation does not reach the detector: (1) scattering
and (2) absorption.  For samples of NP, absorption
strongly dominates.  The intensity of Rayleigh
scattering from a single NP scales as the diameter
to the sixth power, and scales as frequency to the
fourth power.\cite{tanner75}

\section{FIR absorption of metal NP}

Although the focus of this paper is on FIR absorption
of insulating NP, it is instructive to review the
mechanisms of FIR absorption in metal NP.  As noted
above, effects of scattering are negligible.

When the diameters of a NP is much smaller than the
wavelength of the incident radiation, the NP can be
considered to be a dielectric sphere in a static
uniform electric field.  The dielectric response
of a metal NP is related to its conductivity.
The motion of polarization charge corresponds to
current flow in the NP, and ohmic losses lead to
energy absorption.  This is the electric dipole
mechanism.  There is a magnetic dipole mechanism
which starts to become dominant over the electric
dipole mechanism when the NP is larger.
This latter mechanism is also more important when
the conductivity is greater.\cite{tanner75}

The great interest in the subject of FIR absorption
of metal NP is in part due to the fact that the
observed FIR absorption is orders of magnitude
greater than that predicted by the above two
mechanisms.\cite{tanner75,granqvist76}
This has led to a great deal of theoretical
investigations into mechanisms by which the amount
of FIR absorption could be significantly increased.

FIR absorption can be greatly increased by the
presence of an oxide surface layer.\cite{simanek77}
This would have lower conductivity than the metal
interior and provide a mechanism for increased energy
dissipation.  Non-spherical shape of the NP also
increases FIR absorption.\cite{ruppin79,sen82}
Clustering of the NP could be modelled in terms
of percolation\cite{curtin85,noh88} which means
that long conducting paths form.
The electromagnetic response of NP pairs and chains
has also been considered.\cite{fu95}  Clustering
effects can be analyzed in terms of an effective
medium either using the Maxwell-Garnet theory
or other formulas.\cite{marquardt89}  Quantum
mechanical effects have been considered using
the random phase approximation\cite{wood82}
and density functional theory\cite{kurkina96}
and other approaches.\cite{dasgupta81,monreal85}
There may be an enveloping electron cloud in a
classical description.\cite{plyukhin99}

Another mechanism is energy dissipation through
the excitation of phonons.  The excitation of
ultrasound phonons through an electromagnetic wave
hitting bulk metal has long been known.\cite{chimenti74}
This possibility was pointed out for FIR radiation
striking metal NP.\cite{glick78,sheng85,hua85}

\section{Confined Acoustic Phonons}

The focus of this paper is mechanisms of FIR absorption
through coupling of the incident electromagnetic radiation
to the mechanical vibrational modes of the NP.  This
section provides background information on these
acoustic phonons.

Approximations about a NP are made as follows:
It is an isolated sphere of radius $R_p$.
(Our samples had $2R_p$ = 4.5~nm and 7.5~nm.)
It is a homogeneous isotropic elastic continuum.  It has
density $\rho$ and longitudinal and transverse speeds
of sound $v_L$ and $v_T$.  For polycrystalline TiO$_2$,
$\rho$ = 4097~kg/m$^3$, $v_L$ = 8610~m/s and
$v_T$ = 5160~m/s.\cite{minnear77}
These values are for polycrystalline rutile phase TiO$_2$.
The NP in the sample are crystalline and in the anatase phase,
whose elastic properties are expected to be somewhat different.
We do not consider the distribution of NP sizes
within the actual samples or deviations from spherical shape.

Points (material coordinates) within the NP are denoted by
$\vec{R}$.  Spatial position is denoted by $\vec{r}$.  If
the NP is at rest, then $\vec{R} = \vec{r}$
at all points within the NP.  Otherwise, vibrations of
the NP are described in terms of the displacement field
$\vec{u} = \vec{r} - \vec{R}$. $\vec{u}$ is a
function of material coordinate $\vec{R}$ and time, $t$.

In general, since it is a vector field, $\vec{u}$ is the
sum of a zero curl vector field and a zero divergence
vector field.  The zero curl vector field is expressible
as the gradient of a scalar field $\Phi$.  The zero
divergence field is expressible as the curl of
a vector field $\Xi$.
\begin{equation}
\vec{u}  = \nabla \Phi + \nabla \times \Xi
\end{equation}
In turn, $\Xi$ may be expressed in terms of two
scalar fields $\psi$ and $\chi$.
\begin{equation}
\Xi = \nabla \times ( \vec{R} \psi ) +
\nabla \times \nabla \times ( \vec{R} \chi )
\end{equation}
For a normal mode of frequency $\omega$,
the elastic equations of motion of the NP turn
into three independent scalar differential equations:
\begin{eqnarray}
 \nabla^{2} \Phi & = & -\left(\frac\omega{v_L}\right)^{2} \Phi \\
 \nabla^{2} \chi & = & -\left(\frac\omega{v_T}\right)^{2} \chi \\
 \nabla^{2} \psi & = & -\left(\frac\omega{v_T}\right)^{2} \psi
\end{eqnarray}
In spherical material coordinates $R$, $\theta$ and $\phi$, the
above equations have solutions of the following general form
(the $exp(i \omega t)$ factors being omitted):
\begin{eqnarray}
 \Phi & = & j_{\ell}(k_L R) Y_{\ell m} (\theta,\phi) \\
 \chi & = & j_{\ell}(k_T R) Y_{\ell m} (\theta,\phi) \\
 \psi & = & j_{\ell}(k_T R) Y_{\ell m} (\theta,\phi)
\end{eqnarray}
where $j_{\ell}$ are spherical Bessel functions of
the first kind (nonsingular at the origin) of order $\ell$
and $Y_{\ell m}(\theta,\phi)$ the spherical harmonics.
$\ell \geq 0$ is the angular momentum index and
$m$ is the z-component of angular momentum, so
that $-\ell \leq m \leq \ell$.

The boundary condition at $R = R_p$ is that the
$r r$, $r \theta$ and $r \phi$ components of the
stress tensor are all zero.

There are two classes of solution.\cite{lamb1882}  In the first
class (torsional modes, TOR) only $\chi$ is nonzero and
the displacement field $\vec{u}$ has zero divergence.
In the second class (spheroidal modes, SPH) both
$\Phi$ and $\psi$ are nonzero.  There is a special
case when $\ell = 0$ where the motion is purely
radial.  Otherwise, spheroidal motions have both
nonzero curl and divergence.

Modes are also indexed in order of increasing
frequency by integer $n \geq 0$.  Let $p$ denote
TOR or SPH.  Then we indicate a given mode by
($p$,$\ell$,$m$,$n$).  Its frequency is denoted
as $\omega_{p \ell n}$, noting that frequencies
are independent of $m$.  The frequencies of selected
modes of a sphere of TiO$_2$ are given in Tab.~\ref{fsm}.

\begin{table}
\caption{\label{fsm}
Frequencies (cm$^{-1}$) of the IR active (SPH,$\ell=1$,$m$,$n$) modes for  
the HT-3 and HT-7 samples using sizes determined from LFR experiments.  
In addition, the dimensionless frequency $\eta$ for each mode is given, 
where $\eta = \omega R_p / v_T$.                                        
}
\begin{tabular}{|c|c|c|c|c|c|c|c|c|}
\toprule
 $n$ & $\eta$ & $\nu_{\rm HT-3}$ & $\nu_{\rm HT-7}$\\
\colrule
  0  & 3.357  &      40.87       &       24.52     \\
  1  & 6.555  &      79.79       &       47.87     \\
  2  & 7.660  &      93.25       &       55.95     \\
\botrule
\end{tabular}
\end{table}

\begin{figure}
\includegraphics[width=\columnwidth]{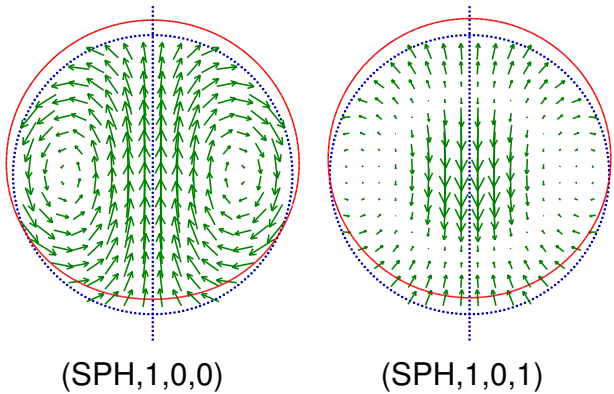}
\caption{\label{fig4}
Acoustic phonon displacement fields within the NP for
two modes are shown.  The vertical blue dotted line is the $z$-axis.
The circular blue dotted line is the equilibrium position of the NP
surface.  The circular red solid line is the displaced profile
of the NP surface.  The magnitude of the displacement is arbitrary.
The (SPH,1,$m$,$n$) modes for $m \neq 0$ are identical to these,
except rotated to have their symmetry axis along $x$ ($m=1$) or
$y$ ($m=-1$).
}
\end{figure}

Figure~\ref{fig4} shows the displacement field $\vec{u}$
for the (SPH,1,0,0) and (SPH,1,0,1) modes.

\section{Selection Rules}

Group theoretical arguments\cite{duval92} may be
applied to the situation of a NP which is (1)
spherically symmetric (2) isolated (3) much smaller
than the wavelength of incident radiation (4) isotropic
in its dielectric and elastic response.  In this case,
only the SPH $\ell=1$ modes can absorb FIR to first
order in the incident intensity.  Note that this
result does not apply to higher
order processes in which the scattered intensity is non-linear
in the incident intensity.

\section{elastic dielectric continuum}

The point of this section is that there is no first order
absorption of FIR by acoustic phonons in a NP modelled as
an elastic dielectric continuum.  What is needed is
consideration of quantum size effects, treated phenomenologically
in a later section.

A continuum description of bulk matter ignores fluctuations
of physical quantities on the distance scale of the crystal
lattice.  Quantities such as density, electric field, charge
density, stress and strain are defined as averages over
regions which are small and yet large compared to the
atomic length scale.

In particular, neutral bulk matter (with no external field) has zero
charge density in this macroscopic description, even though
there are large microscopic fluctuations of the distribution
of charge, more positive where nuclei are located and more
negative elsewhere.

Consequently, a weak electric field does not exert force
on neutral matter, in this description.  If an electric
field is externally applied to a NP, there will be a
polarization field
$\vec{P}(\vec{r})$ within the NP,
leading to bound surface charge density $\sigma_b(\vec{r})
= \vec{P} \cdot \hat{n}$ at the NP surface. $\hat{n}$ is the unit
surface normal vector.
Since $\sigma_b$ is proportional to the applied
field, and since the force on a given point within the NP
is the product of the electric field and the charge density,
therefore the force on the NP is proportional to the square
of the applied field.  Finally, the absorbed power would scale
as the applied field to the fourth power.  Since incident
energy flux goes as field to the second power, there is
zero absorption cross section at lowest order in the
incident electric field.

Direct energy absorption by phonons requires a nonlocal relation
between electron current and field, and also that the
electron mean free path be longer than the penetration
depth of the field, so that the screening is
incomplete.\cite{glick78}

\section{Dipolar Phonon}

In this section, we consider absorption by the
(SPH,$\ell$=1,$m$,$n$) modes.  As mentioned above, group
theoretical symmetry arguments show that the selection
rules allow FIR absorption by these modes.

The mechanism for (SPH,$\ell$=1,$m$,$n$) modes to
absorb FIR radiation requires a description of the NP
that recognizes the different role played by positive
and negative charges within the NP.  It is well known
from density functional theoretical calculations of
metal NP that the electrons in a NP tend
to move outwards relative to the positive nuclei,
leading to a shell of negative charge surrounding
the NP, and with its interior slightly positive.

The simplifying assumption that we employ here is that
the NP has such an outer shell of negative charge,
with total charge $Q$.  The interior of the NP is
positively charged, so that the NP as a whole is
neutral.

\begin{figure}
\includegraphics[width=\columnwidth]{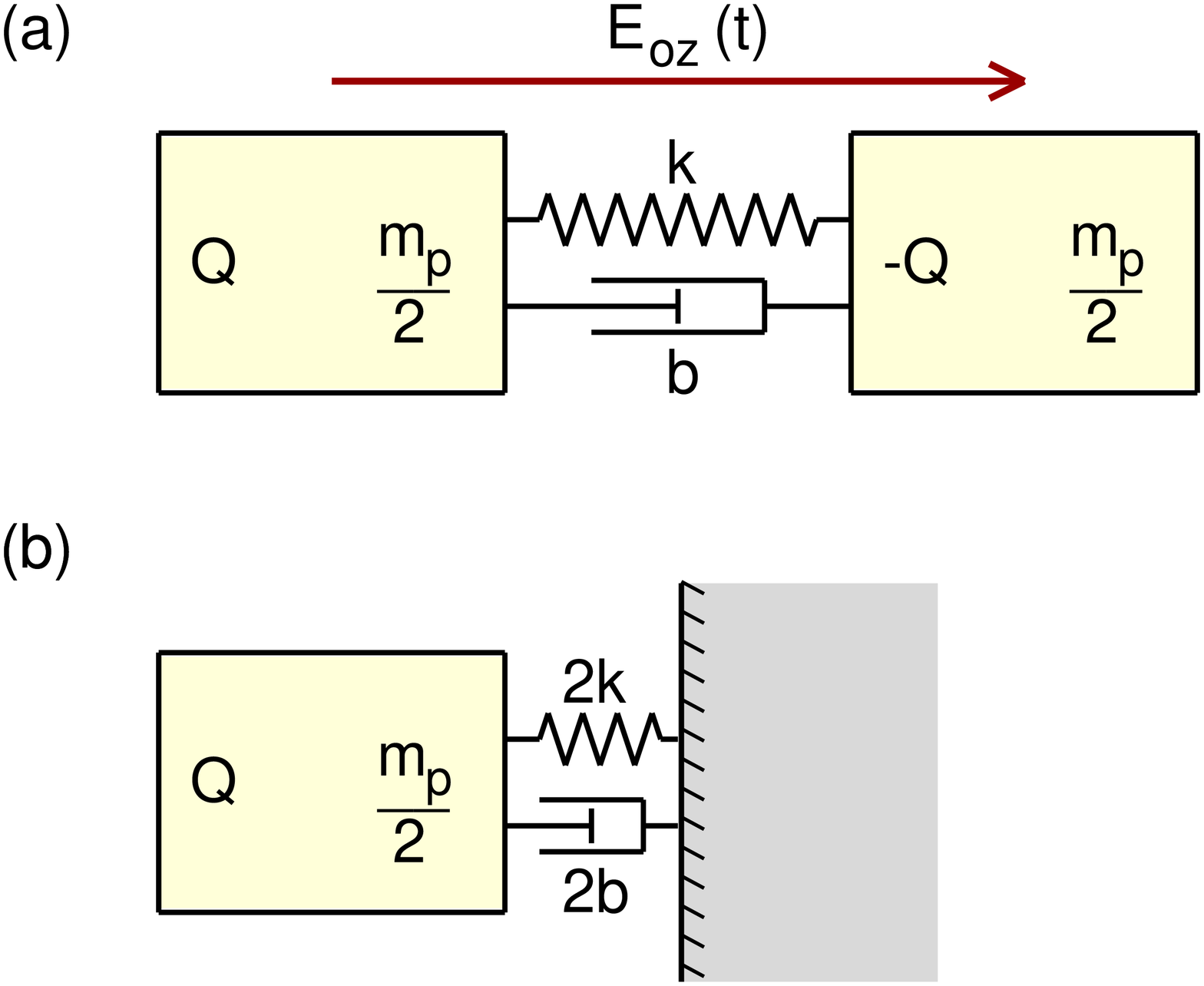}
\caption{\label{fig5}
(a) Harmonic oscillator model of absorption of FIR by internal
vibrations of a NP.  Two masses each of mass $\frac12 m_p$ are
connected by a spring, $k$, and a dashpot, $b$.
(b) Simplified version of (a).  The average power absorbed
by (b) must be doubled to get the average power absorbed by
(a).}
\end{figure}

\section{mass spring model}

We present here a very simplified phenomenological model of
FIR absorption due to the internal vibration of a NP.  Let
the acoustic phonon (we are thinking of the (SPH,1,0,$n$)
mode for some $n$) be represented by dividing the NP into
two pieces, each of mass $m_p / 2$, as shown in Fig.~\ref{fig5}(a).
The two parts are connected by a spring with force constant
$k$.  With only this spring, this system has natural
frequency $2 \sqrt{k/m_p}$.  Furthermore, the system is damped
by means of a dashpot connecting the two masses.  A dashpot
(or shock absorber) exerts a resistance force $F = b v$
where $v$ is the speed.  The physical origin of the damping
is not important, but transfer of mechanical energy through
physical contact with neighbouring NP in the nanopowder
would be one important mechanism.

By symmetry, Fig.~\ref{fig5}(a)
can be simplified to Fig.~\ref{fig5}(b).  However, when
finding the total power absorption in the NP, the
final answer will need to be doubled.

First, considering free damped vibrations of this system,
let $z(t)$ denote the displacement of the block in
Fig.~\ref{fig5}(b).  The equation of motion is
\begin{equation}
-2k z - 2b \frac{dz}{dt} = \frac{m_p}{2} \frac{d^2 z}{dt^2}
\end{equation}
Assuming $z(t) = exp(i \omega t)$
\begin{equation}
-2k - 2 i \omega b  = -\frac12 m_p \omega^2
\end{equation}

Solving this quadratic, the frequency of free oscillations
is a complex number with imaginary part
\begin{equation}
\gamma = 2b / m_p
\label{eq_gamma}
\end{equation}

Next, considering steady state driven vibrations of this
system,
let the applied electric field (the incident FIR radiation)
be $E_{oz}(t) = E_{oz} exp(i \omega t)$.
The equation of motion of the system is
\begin{equation}
Q E_{oz}(t) - 2k z(t) - 2b \frac{dz}{dt} = \frac12 m_p \frac{d^2 z}{dt^2}
\end{equation}

The displacement of the block in Fig.~\ref{fig5}(b)
is $z(t) = B \, exp(i \omega t)$ where the amplitude
$B$ is to be determined below.  The equation of motion becomes
\begin{equation}
Q E_{oz} - 2k B - 2 i \omega b B = - \frac12 m_p \omega^2 B
\end{equation}
so that
\begin{equation}
B = \frac{Q E_{oz}}{2 k + 2 i b \omega - \frac12 m \omega^2}
\end{equation}

At resonance, the real part of the denominator vanishes,
so that
\begin{equation}
B = \frac{Q E_{oz}}{2 \, i \, b \, \omega}
\end{equation}

The velocity of the block is $v_z = dz/dt = v_{oz} exp(i \omega t)$
\begin{equation}
v_{oz} = \frac{Q E_{oz}}{2 b}
\end{equation}

The instantaneous dissipated power $P(t)$ is the applied force
on the block, $F_z(t) = Q E_{oz}(t)$, times the velocity $v_z(t)$.  This is
integrated over one full cycle to yield the average dissipated
power.  Finally, the result is multiplied by 2 to regain
the situation of Fig.~\ref{fig5}(a).
\begin{equation}
P = \frac{(Q E_{oz})^2}{2 b}
\end{equation}

From considering the non-driven but damped oscillations, we saw
in Eq.~(\ref{eq_gamma}) that $b = \frac12 \gamma m_p$, where $\gamma$ is the imaginary
part of the frequency of damped oscillations.
\begin{equation}
P = \frac{(Q E_{oz})^2}{\gamma m_p}
\end{equation}

The incident power flux is given by $S$ = $\frac12 c \, \epsilon_o E_{oz}^2$
where $E_{oz}$ is the peak amplitude of the incident electric
field.  $c$ is the speed of light in vacuum and $\epsilon_o$ is
the permittivity of free space.  
For simplicity, the dielectric constant of the matrix in which
the NP is supported is assumed to be $\epsilon_o$.
The excess absorption cross section for a single NP is found
from $\sigma_{ex} = P / S$, yielding
\begin{equation}
\label{eq_sigma_th}
\sigma_{ex} = \frac{2 Q^2}{\gamma \epsilon_o c \, m_p}
\end{equation}

We call this the excess since there are other absorption
mechanisms for FIR not related to acoustic phonons.

\section{Samples}

Anatase TiO$_2$ nanopowders have been prepared by hydrothermal
synthesis in subcritical water. Synthesis conditions have
already been described elsewhere.\cite{MillotJECS05,PighiniJNR06}
Two powders with different average NP size have been
used in this work. They were characterized by X-ray diffraction,
surface area measurements, high-resolution electron microscopy
and Raman spectroscopy, including LFR, in a previous work\cite{PighiniJNR06}
where they are labelled HT-3 and HT-7.  These techniques were
used to determine the crystalline structure and the size
of the NP.  The average NP diameter, as estimated from LFR,
is roughly 4.5~nm for HT-3 and 7.5~nm for HT-7, slightly different
values being obtained depending on the technique.
LFR revealed well-resolved peaks contrary to
previous works on similar nanopowders. This is the sign of a
relatively narrow size distribution which is necessary to
study size-dependent properties.

\section{Infrared Measurement}

The nanopowder samples were adhered to adhesive tape.
FIR absorption measurements were made in a vacuum chamber.
The room-temperature transmittance measurements were
performed using a Bruker IFS 66v Fourier transform infrared
spectrometer for the 15 - 200~cm$^{-1}$ frequency range with a 4.2~K
silicon bolometer. The transmittance ratio of the nanopowders
with tape to the tape alone was obtained.

The circular aperture through which the IR passes is about
5~mm in diameter so that the aperture area is $A$ =
2.0$\times$10$^{-5}$~m$^{2}$.

We did not do any heat treatments of the TiO$_2$ powder before
the IR measurements. Adsorption measurements have shown that
the surfaces of the individual NP in these
powders is covered by adsorbed water.
Applying a vacuum without heating results in a weight loss of
approximately 10\%. Subsequent
heating up to 250$^o$C results in an additional loss of 3\%.       
Condensed water has a weak absorption in the low-frequency range which  
has been assigned to the ``librational modes of structural aggregates    
of clustered molecules''.\cite{HastedCPL85,VijCPL89,DuvalJCP90} No       
such signal is expected in the present experimental results because     
the small amount of water is adsorbed on a large surface ($\sim$~300~m$^2$/g)
so no such aggregates are possible.                                               

We have tried using high density polyethylene (HDPE) as
the substrate, as an alternative to the adhesive tape.  HDPE is
known to be nearly transparent in the FIR region.   We found
that HDPE shows somewhat higher transmitted intensity than
that of the adhesive tape.  However, it is difficult to affix
the TiO$_2$ powders on the HDPE surface.

The properties of the two samples are shown
in Tab.~\ref{prop}.  The mass of one NP, $m_{p}$, is
found from $m_{p} = \frac43 \pi \rho R_p^3$.

\begin{table}
\caption{\label{prop}
Properties of the two TiO$_2$ nanopowder samples are given.
The sample designations HT-3 and HT-7 are as used in the reports
of the nanopowder synthesis.\cite{PighiniJNR06}
The NP mass is determined assuming
a density of 4097~kg/m$^3$.\cite{minnear77}
}
\begin{tabular}{|c|c|c|c|}
\toprule
       & sample designation                 & HT-3  & HT-7\\
\hline
$2R_p$ & diameter from LFR (nm)             & 4.5   & 7.5\\
$m_p$  & mass of a single NP ($10^{-22}$kg) & 1.95  & 9.0\\
       & mass of NP on tape (mg)            & 5.6   & 4.2\\
$N$    & number of NP on tape ($10^{15}$)   & 29    & 4.6\\
\botrule
\end{tabular}
\end{table}

\section{Transmittance Spectra}

\begin{figure}
\includegraphics[width=\columnwidth]{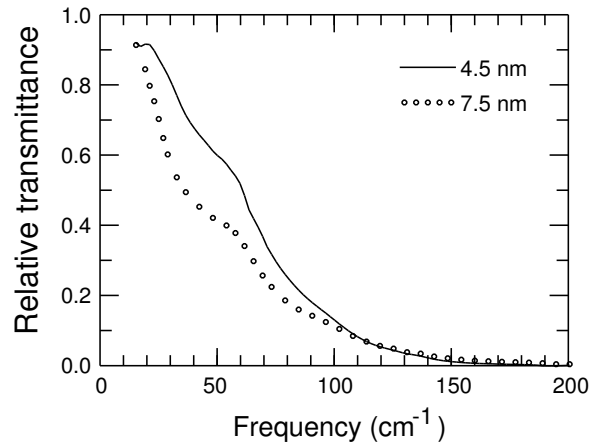}
\caption{\label{fig1}
Transmitted intensity ratio versus frequency for two
TiO$_2$ nanopowders. The solid line is 4.5 nm~diameter nanopowder.
The circles are for 7.5~nm diameter nanopowder.
The vertical
scale is the ratio of transmittance with and without
nanopowder affixed to the adhesive tape.}
\end{figure}

FIR transmission spectra are separately collected for
(1) adhesive tape only ($I_{T0}(\omega)$) and (2) adhesive tape
with TiO$_2$ nanopowder adhering ($I_{T}(\omega)$).
Necessarily, $I_{T}(\omega) \leq I_{T0}(\omega)$.
Figure~\ref{fig1} shows the transmittance
intensity ratio spectra ($I_{T}/I_{T0}$)
of the two samples. By definition, the
transmittance intensity ratio cannot exceed 1.

Using Eq.~(\ref{eq_sigma_exp}) it is possible to determine the
experimental absorption cross section per NP.  As in the case of FIR
absorption by metal NP, $\sigma(\omega)$ is
generally of the form $C \omega^2$ where $C$ is a constant.
Let the excess absorption cross section, $\sigma_{ex}(\omega)$
be defined by
\begin{equation}
  \sigma_{ex}(\omega) = - \frac{A}{N} \ln\left(\frac{I_T}{I_{T0}}\right)
                    - C \omega^2
\end{equation}
where $C$ is the optimum value of the constant to remove
the quadratic behavior.  $\sigma_{ex}(\omega)$ is the
portion of the absorption cross section due to mechanisms
that only apply at special frequencies.

\begin{figure}
\includegraphics[width=\columnwidth]{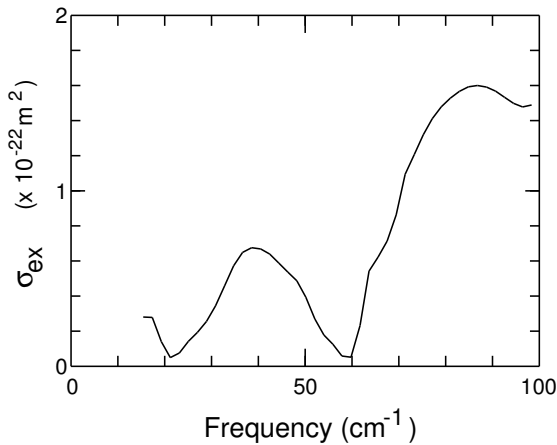}
\caption{\label{fig2}
For the 4.5~nm diameter NP sample,
excess absorption cross section per NP is plotted
versus the frequency in wavenumbers.}
\end{figure}

\begin{figure}
\includegraphics[width=\columnwidth]{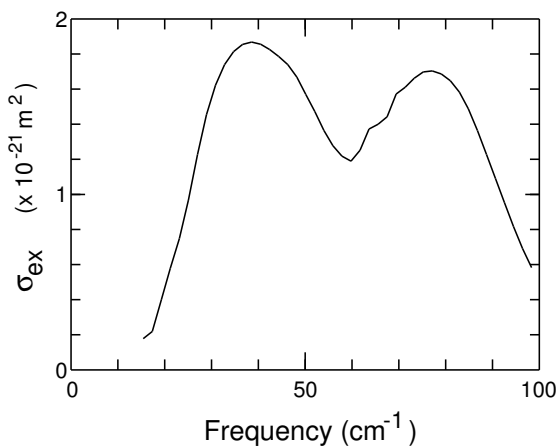}
\caption{\label{fig3}
For the 7.5~nm diameter NP sample,
excess absorption cross section per NP is plotted
versus the frequency in wavenumbers.}
\end{figure}

In Fig.~\ref{fig2} and Fig.~\ref{fig3} a quadratic background
of the form $C \omega^2$ has been subtracted from the absorption
cross section per NP.  This reveals peaks.  For each sample,
there is a peak at approximately 40~cm$^{-1}$.
It should be noted that while the fit of the quadratic
background is straightforward for the HT-3 sample, it does not
work so well at higher frequencies for the HT-7 sample.  However,
the resulting peak position in $\sigma_{ex}(\omega)$
is not sensitive
to this uncertainty in fitting the background.

\section{Discussion}

As Fig.~\ref{fig2} and Fig.~\ref{fig3} illustrate, there
are peaks in the absorption cross section
once a quadratic background is subtracted.
The two peaks at approximately 40~cm$^{-1}$ can be
identified with the 40.87~cm$^{-1}$ (SPH,1,0,0) mode
for the 4.5~nm diameter sample and with the
47.87~cm$^{-1}$ (SPH,1,0,1) mode for the 7.5~nm
diameter sample.
From these two figures, the corresponding peak value of
$\sigma_{ex}$, the excess absorption cross section per NP,
is given in Tab.~\ref{tabres}.

\begin{table}
\caption{\label{tabres}
The values for $\sigma_{ex}$ are the peak heights
from Fig.~\ref{fig2} and Fig.~\ref{fig3}.  The upper
bound for $Q$ is from Eq.~(\ref{eq_Q}).  The number
of TiO$_2$ groups in each NP is indicated.
The surface area of each NP in nm$^2$ is given,
along with the number of adsorbed H$_2$O molecules,
assuming 3 H$_2$O per square nanometer.
}
\begin{tabular}{|c|c|c|c|}
\toprule
              &          & HT-3                & HT-7                  \\
\hline
$\sigma_{ex}$ & m$^2$    & 7$\times$10$^{-23}$ & 1.9$\times$10$^{-21}$ \\
 $Q$          & Coulombs & $40 \, e$              & $350 \, e$    \\
TiO$_2$       & groups   & 1460                & 6770       \\
$4\pi R_p^2$ & nm$^2$   & 64         & 180        \\
H$_2$O   & adsorbed &  190      & 540        \\
\botrule
\end{tabular}
\end{table}

In a realistic environment such as a nanopowder, the
NP acoustic phonon modes would be damped by mechanisms
such as contact with neighbouring NPs.  In this case,
the frequency of the damped acoustic phonon can be
treated as a complex frequency $\omega + i \gamma$,
where the imaginary part $\gamma$ corresponds to damping
of the vibration amplitude with time of the form
$exp(-\gamma t)$.  The value of $\gamma$
can be bounded above based on LFR peak
linewidths.\cite{PighiniJNR06}
However, LFR linewidths are due
to two phenomena: (1) NP size variation in the sample and
(2) acoustic phonon damping.  In other words, it is
conceivable that the acoustic phonons are very lightly
damped and that the observed LFR peak width is due to NP
size variation alone.  A rough estimate for the bound is that
$\gamma < 0.3 \, \omega$.

Using Eq.~(\ref{eq_sigma_th}), $Q^2$ may be bounded above:
\begin{equation}
\label{eq_Q}
Q^2 < \frac12 0.3 \, \omega \, \sigma_{ex} \, \epsilon_o \, c \, m_p
\end{equation}

The values of $Q$ resulting from this relation are shown
in Tab.~\ref{tabres}.  Since $\omega = \eta v_T / R_p$
and $m_p$ = $\frac43 \pi \rho R_p^3$, this is equivalent to
\begin{equation}
Q^2 < \frac23 0.3 \, \pi \, \eta \, v_T \, \sigma_{ex} \, \epsilon_o \, c \, \rho \, R_p^2
\end{equation}

The molecular weight of TiO$_2$ is 79.9 g/mol, so that a
single ``molecule'' of TiO$_2$ has mass $M$ = 1.33$\times$10$^{-25}$~kg.
The number of TiO$_2$ groups for NP in each sample is
shown in Tab.~\ref{tabres}.

The volume occupied by each TiO$_2$ group is $M/\rho$.
Assuming that each TiO$_2$ group contributes one electron,
the volume of the corresponding region is $M Q / (\rho e)$.
Dividing this by $4 \pi R_p^2$ yields
the thickness of a surface layer on the NP in which the
required charge $Q$ is found.  Since $Q$ is bounded above,
this is an upper bound on the layer thickness.
For the 4.5~nm and 7.5~nm samples, respectively, the maximum
required charge layer thicknesses are 2.1$\times$10$^{-11}$~m
and 6.4$\times$10$^{-11}$~m.  These layer thicknesses are
on the order of the charge screening length inside bulk
material.

Alternatively, we note that the remaining adsorbed water
molecules on the NP surface have a density roughly
estimated as 3 molecules
per square nanometer.  Table~\ref{tabres} shows the number
of H$_2$O molecules for each sample.  It is conceivable
that the charge on these strongly polar adsorbed molecules
might play a role in the coupling of the
(SPH,$\ell$=1,$m$,$n$) modes to FIR.

\section{Conclusion}

This study presents the first direct evidence of absorption
of FIR electromagnetic radiation by acoustic
phonon modes of NP.  The remarkably close
agreement between the observed positions of the peaks
(in Fig.~\ref{fig2} and Fig.~\ref{fig3}) and the
theoretical estimates of the frequencies of the
(SPH,$\ell$=1,$m$,$n$) acoustic phonon modes provides
evidence for the correctness of this interpretation.
In addition, the excess absorption cross section
(above the smooth quadratic background) per NP is in
plausible agreement with a simple theoretical model.

Additional studies will be needed to verify the agreement in
observed and predicted peak position over a wider range of NP
dimensions and materials. $\sigma_{ex}$ needs to be measured
for nanopowder samples with varying $R_p$, NP material,
and conditions affecting adsorbed surface materials.
This could provide an experimental test for any proposed
improvements to the simple theoretical model of NP FIR
absorption that we have proposed.

Once more extensively studied, FIR absorption by NP acoustic
phonons has the potential to become an important diagnostic
probe of nanopowder properties, such as particle size
distribution, complementing existing techniques such as LFR
and UFLPP.  In particular, FIR absorption is sensitive to the
(SPH,$\ell$=1,$m$,$n$) modes while LFR is primarily
sensitive to the (SPH,$\ell$=2,$m$,0) modes
and UFLPP is sensitive only to the (SPH,0,0,$n$) modes.
Comparison of frequencies for these three kinds of
modes could be a useful test of the accuracy of the
elastic continuum model of NP vibrations.

\section{Acknowledgements}

D.~B.~M. acknowledges support from the Natural Sciences and
Engineering Research Council of Canada.
H.~L.~L. acknowledges support from the National Science Council
of the Republic of China under grants NSC 94-2112-M-003-002
and 94-2120-M-007-013, and from National Taiwan Normal University
under Grant No. ORD93-B.
M.~Blades and A.~Plyukhin are thanked for their assistance.

\providecommand{\href}[2]{#2}\begingroup\raggedright\endgroup


\begin{thebibliography}{10}

\bibitem{gorkov65}
L.~P. Gorkov and G.~M. Eliashberg {\em Sov. Phys.-JETP} {\bf 21} (1965) 940.

\bibitem{halperin86}
W.~P. Halperin {\em Rev. Mod. Phys.} {\bf 58} (1986) 533.

\bibitem{tanner75}
D.~B. Tanner, A.~J. Sievers, and R.~A. Burhman {\em Phys. Rev. B} {\bf 11}
  (1975) 1330.

\bibitem{granqvist76}
C.~G. Granqvist, R.~A. Buhrman, J.~Wyns, and A.~J. Sievers {\em Phys. Rev.
  Lett.} {\bf 37} (1976) 625.

\bibitem{devaty84}
R.~P. Devaty and A.~J. Sievers {\em Phys. Rev. Lett.} {\bf 52} (1984) 1344.

\bibitem{lee85}
S.~I. Lee, T.~W. Noh, K.~Cummings, and J.~Gaines {\em Phys. Rev. Lett.} {\bf
  55} (1985) 1626.

\bibitem{devaty90}
R.~P. Devaty and A.~J. Sievers {\em Phys. Rev. B} {\bf 41} (1990) 7421.

\bibitem{helle05b}
M.~Helle, A.~Harju, and R.~M. Nieminen {\em Phys. Rev. B} {\bf 72} (2005)
  205329.

\bibitem{marlo03}
M.~Marlo, A.~Harju, and R.~M. Nieminen {\em Phys. Rev. Lett.} {\bf 91} (2003)
  187401.

\bibitem{marlohelle05}
M.~{Marlo-Helle}, A.~Harju, and R.~M. Nieminen {\em Physica E} {\bf 26} (2005)
  286.

\bibitem{simanek77}
E.~Sim{\'a}nek {\em Phys. Rev. Lett.} {\bf 38} (1977) 1161.

\bibitem{ruppin79}
R.~Ruppin {\em Phys. Rev. B} {\bf 19} (1979) 1318.

\bibitem{sen82}
P.~N. Sen and D.~B. Tanner {\em Phys. Rev. B} {\bf 26} (1982) 3582.

\bibitem{curtin85}
N.~W. Curtin and N.~W. Ashcroft {\em Phys. Rev. B} {\bf 31} (1985) 3287.

\bibitem{noh88}
T.~W. Noh, S.~I. Lee, Y.~Song, and J.~R. Gaines {\em J. Phys. C: Solid State
  Physics} {\bf 21} (1988) 1849.

\bibitem{fu95}
L.~Fu and L.~Resca {\em Phys. Rev. B} {\bf 52} (1995) 10815.

\bibitem{marquardt89}
P.~Marquardt and G.~Nimtz {\em Phys. Rev. B} {\bf 40} (1989) 7996.

\bibitem{wood82}
D.~M. Wood and N.~W. Ashcroft {\em Phys. Rev. B} {\bf 25} (1982) 6255.

\bibitem{kurkina96}
L.~I. Kurkina and O.~V. Farberovich {\em Phys. Rev. B} {\bf 54} (1996) 14791.

\bibitem{dasgupta81}
B.~B. Dasgupta {\em Phys. Rev. B} {\bf 24} (1981) 554.

\bibitem{monreal85}
R.~Monreal, P.~{de Andr{\'e}s}, and F.~Flores {\em J. Phys. C: Solid State
  Phys.} {\bf 18} (1985) 4951.

\bibitem{plyukhin99}
A.~V. Plyukhin, A.~K. Sarychev, and A.~M. Dykhne {\em Phys. Rev. B} {\bf 59}
  (1999) 1685.

\bibitem{chimenti74}
D.~E. Chimenti {\em Phys. Rev. B} {\bf 10} (1974) 3228.

\bibitem{glick78}
A.~J. Glick and E.~D. Yorke {\em Phys. Rev. B} {\bf 18} (1978) 2490.

\bibitem{sheng85}
P.~Sheng {\em Phys. Rev. B} {\bf 31} (1985) 4906.

\bibitem{hua85}
X.~M. Hua and J.~I. Gersten {\em Phys. Rev. B} {\bf 31} (1985) 855.

\bibitem{minnear77}
W.~P. Minnear and R.~C. Bradt, ``Elastic properties of polycrystalline
  {TiO$_{2-x}$},'' {\em J. Am. Ceramic Soc.} {\bf 60} (1977) 458--459.

\bibitem{lamb1882}
H.~Lamb {\em Proc. London Math. Soc.} {\bf 13} (1881-1882) 189--212.

\bibitem{duval92}
E.~Duval {\em Phys. Rev. B} {\bf 46} (1992) R5795--R5797.

\bibitem{MillotJECS05}
N.~Millot, B.~Xin, C.~Pighini, and D.~Aymes, ``Hydrothermal synthesis of
  nanostructured inorganic powders by a continuous process under supercritical
  conditions,'' {\em J. Europ. Ceram. Soc.} {\bf 25} (2005) 2013.

\bibitem{PighiniJNR06}
C.~Pighini, D.~Aymes, N.~Millot, and L.~Saviot, ``Low-frequency {R}aman
  characterization of size-controlled anatase {TiO$_2$} nanopowders grown by
  continuous hydrothermal synthesis,'' {\em J. Nanoparticle Research}.
  accepted.

\bibitem{HastedCPL85}
J.~B. Hasted, S.~K. Husain, F.~A.~M. Frescura, and J.~R. Birch, ``Far-infrared
  absorption in liquid water,'' {\em Chem. Phys. Lett.} {\bf 118} (1985)
  622--625.

\bibitem{VijCPL89}
J.~K. Vij, ``Millimeter and submillimeter laser spectroscopy of water,'' {\em
  Chem. Phys. Lett.} {\bf 155} (1989) 153--156.

\bibitem{DuvalJCP90}
J.~L. Rousset, E.~Duval, and A.~Boukenter, ``Dynamical structure of water:
  Low-frequency {R}aman scattering from a disordered network and aggregates,''
  {\em J. Chem. Phys.} {\bf 92} (1990) 2150--2154.

\end{thebibliography}
\end{document}